
\input harvmac      

\def\a{\alpha} \def\b{\beta}  \def\G{\Gamma}
\def\d{\delta}  \def\ee{\varepsilon} 
   
 \def\l{\lambda} \def\L{\Lambda}  
  \def\p{\pi}  \def\r{\rho}
\def\s{\sigma}   
    
\def\Ps{\Psi} \def\o{\omega}  
\def\pa{\partial}   \def\half{{1\over
2}} \def\dx{d^3\!x\ }\def\dy{d^3\!y\ }\def\IR{I\!\!R}

\Title{}{Gauge Invariant Geometric Variables
For Yang-Mills Theory\footnote{$^*$}
{This work is supported in part by funds provided by the U.S.
Department of Energy (D.O.E.) under cooperative agreement
\#DE-FC02-94ER40818 and by NSERC of Canada and FCAR of Qu\'ebec.
E-mail: {\tt
haagense@cinelli.physics.mcgill.ca, knjhnsn@mitlns.mit.edu,
lam@hep.physics.mcgill.ca}.}}

\centerline{Peter E. Haagensen$^{1}$, Kenneth Johnson$^{2}$, and C.S.
Lam$^{1}$} \medskip
\centerline{ ${}^1${\it Physics Department, McGill University}}\vskip-.15cm
\centerline{ {\it 3600 University St.}}\vskip-.15cm
\centerline{ {\it Montr\'eal~~H3A 2T8~~CANADA.}}
\medskip \centerline{ ${}^2${\it Center for
Theoretical Physics, Laboratory for Nuclear Science}}\vskip-.15cm
\centerline{ {\it
Massachusetts Institute of Technology, Cambridge, MA~~02139, USA.}}

\bigskip

\nref\hj{P.~E.  Haagensen and K.~Johnson, {\it Nucl.~Phys.} {\bf B439}
(1995) 597.}
\nref\previous{K.~Johnson, {\it ``The Yang-Mills Ground
State''}, in {\it QCD -- 20 Years Later}, Aachen, June 1992; P.~E.
Haagensen, {\it ``On the Exact Implementation of
Gauss' Law in Yang-Mills Theory"}, in {\it Proceedings of the XIII
Partices and Nuclei International Conference}, Perugia, July 1993 (World
Scientific 1994), {\tt hep-ph/9307319}.}
\nref\fhjl{D.~Z.  Freedman, P.~E.  Haagensen, K.~Johnson
and J.~I.  Latorre, {\it ``The Hidden Spatial Geometry of Non-Abelian
Yang-Mills Theories"}, MIT preprint CTP\#2238, {\tt hep-th/ 9309045}.}
\nref\bfh{M.~Bauer, D.~Z.  Freedman and P.~E. Haagensen, {\it Nucl.~Phys.}
{\bf B428} (1994) 147.}
\nref\mrst{P.~E. Haagensen, {\it ``New Gauge Invariant Variables for
Yang-Mills Theory"}, in {\it Proceedings of the XVII Annual MRST Conference},
Rochester, April 1995, {\tt hep-th/9505188}.}
\nref\fk{T.T.  Wu and C.N.  Yang, {\it Phys. Rev.} {\bf D12} (1975)
3843,3845.  See also D.~Z. Freedman and R.~R. Khuri,  {\it Phys.~Lett.} {\bf
329B} (1994) 263 for a more complete list of references.}
\nref\bps{A. Belavin, A. Polyakov, A. Schwartz and Y. Tyupkin,
{\it Phys.~Lett.} {\bf 59B} (1975) 82.}
\nref\cdg{V. de Alfaro, S. Fubini and G. Furlan, {\it Phys.~Lett.}
{\bf 65B} (1976) 163; C. Callan, R. Dashen and D. Gross,
{\it Phys.~Rev.} {\bf D17} (1978) 2717.}
\nref\man{S. Mandelstam, {\it Phys.~Repts.} {\bf 23} (1976) 245.}
\nref\witten{E. Witten, {\it Nucl.~Phys.} {\bf B311} (1988) 46.}

\centerline{{\bf Abstract}}\medskip

In a previous publication \hj, local gauge invariant geometric variables
were introduced to describe the physical Hilbert space of Yang-Mills
theory.  In these variables, the electric energy
involves the inverse of an operator which can generically have zero
modes, and thus its calculation is subtle.  In the present work, we
resolve these subtleties by considering a small deformation in
the definition of these variables, which in the end is removed.
The case of spherical configurations of the gauge invariant variables is
treated in detail, as well as the inclusion of infinitely heavy point
color sources, and the expression for the associated electric field is
found explicitly. These spherical geometries are seen to correspond
to the spatial components of instanton configurations. The related
geometries
corresponding to Wu-Yang monopoles and merons are also identified.

\vfill

\bigskip
\noindent CTP\#2492

\noindent McGill-95/60

\Date{11/95}

\newsec{Introduction}\bigskip

A new formulation of nonabelian gauge theory has recently been given
\hj\ (hereafter referred to as I.), which is founded on a geometrical
basis and seeks to define a setting in which gauge symmetry is implemented
exactly and manifestly, even under approximations of the dynamics.
Here, we will present further developments of this formulation
as well as an example showing how it may allow one to obtain new
insights into the reasons why the Yang-Mills field theory has a
mass gap and produces a long range confining interaction between
massive colored sources. We have introduced this geometric basis in
the Hamiltonian formulation because in the Lagrangian path integral
formalism for a nonabelian gauge theory, the local gauge invariance
does not really manifest itself as a symmetry,
but rather as a redundancy in the path integral
measure. Any approximations in that formalism are likely to introduce
gauge artifacts precisely because local gauge invariance is not acting
as a symmetry. In the strong coupling region of the theory, an
alternative and perhaps better procedure might be one that avoids such
gauge artifacts. This is possible in the canonical Hamiltonian
formulation, because in ``temporal" gauge, $A_0^a=0$, there is a
remaining local gauge invariance restricted to space-dependent
transformations at a fixed time, and one can achieve the goal of
treating this gauge invariance like a true quantum mechanical symmetry.
The local generators of such gauge transformations form an algebra and
represent symmetries of the Hamiltonian which can be maintained exactly,
even when the dynamics is done approximately. In view of this, one can
understand better the statement that gauge invariance is not a symmetry
away from a fixed-time formalism, and why approximations introduce gauge
artifacts: the {\it full} classical gauge group includes time-dependent
gauge transformations, and these are coupled to the dynamics. An approximate
version of the dynamics is then likely to destroy any attempt to keep
the full gauge invariance.

We implement the gauge symmetry by considering a change of variables in
the Hilbert space such that any function of the transformed variables is
a singlet under the gauge group, {\it i.e.}, is gauge invariant. In fact,
this can be done in different ways  \hj-\mrst. In
I., we have introduced such a transformation of variables, and here
we will develop this formalism in more detail. Our basic procedure
is straightforward and simple to state:  rather than using the space
components of the vector potential, $A_i^a$ ($a$ is a color index,
$i=1,2,3$ a space index), as fundamental coordinates in the Hilbert space
of the theory, we use local quantities which transform covariantly
under gauge transformations.  Whereas the generator of gauge transformations
in terms of $A_i^a$ is complicated by the noncovariant transformation
properties of the vector potential, when expressed in terms of gauge
covariant variables it simply turns into a (color) rotation generator.
Gauge covariant quantities can furthermore be contracted with themselves in
color and lead to gauge invariant variables.  In terms of such gauge
invariant variables, Gauss' law, or gauge symmetry, becomes manifest.

Nevertheless, not any choice of gauge covariant variables is
appropriate.  An appropriate set of variables should describe the correct
number of gauge invariant degrees of freedom at each point of space, and
should also be free of ambiguities (such as, for instance,
Wu-Yang ambiguities, where several gauge unrelated vector potentials
may lead to the same color magnetic field \fhjl,\fk). In I., the set of gauge
covariant variables, $u_i^a$, we have chosen to define is
given by the following differential equations:
\eqn\defi{\ee^{ijk}D_ju_k^a\equiv \ee^{ijk}(\pa_j u_k^a+f^{abc}
A_j^bu_k^c)=0\ .}
The linear operator $\ee^{ijk}D_j\equiv (S^j)^{ik}D_j$, where $S$ is
the single gluon ``spin" operator, plays a central
role in our formulation. When $A_i^a$ is a pure gauge, the
eigenvalues of this operator are $\pm p$ and $0$; in this
case the zero modes are the ``longitudinal" gluons. In general,
the zero mode wavefunctions replace the vector potential
as the dynamical coordinate. Further, the
remaining spectrum and eigenfunctions of the operator enter in
the process of obtaining an expression for the electric field conjugate to
$A_i^a$ in terms of the variables  $u_i^a$.
It is clear that the spectrum is gauge
invariant and that the wavefunctions $u_i^a$ transform as vectors under
gauge changes. In I. and here we actually only consider in detail the
$SU(2)$ theory, $f^{abc}=\ee^{abc}$ (although in I. the extension to
$SU(N)$ is also partially treated).  Because most of the details of
calculation needed for our purposes have been spelled out in I., here we
will rather present a brief summary of previous results.

It turns out that for the $SU(2)$ theory in canonical
formalism, there is a natural symmetry under coordinate
reparametrizations, which is respected by all commutators
and basic formulas except for the Hamiltonian itself.  We have purposefully
maintained this symmetry in defining our new variables, so that a
natural geometric picture arises as a guiding principle in the formalism
at no extra cost.  Under this reparametrization symmetry, the vector
potential transforms as a covariant vector, while both the electric
field $E^{ai}=-i\d/\d A_i^a$ and the magnetic field,
\eqn\bofa{B^{ai} =\ \ee^{ijk}(\pa_j A^a_k+\half\
\ee^{abc} A^b_jA^c_k)\ ,}
transform as contravariant vector densities.
The canonical commutators between coordinates $A_i^a$ and momenta
$E^{bj}$ also transform covariantly, and the Gauss law generator,
\eqn\gauss{{\cal G}^a\ =\ D_i E^{ai}\ \equiv\ \pa_i E^{ai}+
\ee^{abc}A^b_iE^{ci}\ ,}
transforms as a scalar density. It is only because of this
reparametrization covariance that we introduce the above
seemingly peculiar placement of space indices. The only
failure in reparametrization covariance comes in the Hamiltonian:
\eqn\ham{H\ =\ \half\ \int \dx
\left({B^{ai}B^{ai}\over g_s^2}+ g_s^2E^{ai}E^{ai}\right) \ .}
The integrand above is not a geometric scalar with the correct
density weight, and
the contraction in space indices is made with a Kronecker $\d_{ij}$
rather than with a metric tensor, that is, the Hamiltonian is
``committed" to a flat space.

It is straightforward to check that the definition of $u_i^a$ in \defi\ is
identical to the standard geometric equation defining the spin
connection in terms of the Christoffel connection (or vice-versa):
\eqn\covu{\pa_j u^a_k+\ee^{abc}A^b_j u^c_k\
-\G^s_{jk}u^a_s\ =0,}
where $u_i^a$ is a 3-bein, $\ee^{abc}A^b_i$ is a
spin connection, and $\G^i_{jk}$ is the Christoffel connection of the
metric $g_{ij}=u_i^au_j^a$.  Requiring that $u_i^a$ transform as a
vector under both gauge and reparametrization transformations then gives
us a gauge and reparametrization covariant definition, and the above
simple geometric picture.  The ``metric" tensor $g_{ij}=u_i^au_j^a$
neatly organizes the six local gauge invariant degrees of freedom of the
problem into a symmetric $3\times 3$ matrix, and the next task would be
to write the Hamiltonian in terms of these variables.  One can then
prove that any gauge invariant functional of $A_i^a$ can be written as a
function of $g_{ij}$ only, and that any functional of $g_{ij}$ is gauge
invariant (cf.  I.).  This implements gauge invariance exactly. Further,
it is also easy to include other color variables into the formalism
when $u_i^a$ is used as the independent variable.

Before we proceed to write down the gauge invariant, geometric
expressions for the quantities of interest, we shall first observe that
there are zero mode problems associated with the calculation of the
electric field $E^{ai}=-i\d/\d A_i^a$.  It is easy to see that, under
the transformation of variables \defi , the Jacobian matrix $\d A_i^a/\d
u_j^b$ involves the operator $\ee^{ijk}D_j$, which may have more than one
remaining zero mode when the potential is not a pure gauge (where that is of
course $u_i^a$ itself). It then seems there will be an indefinition in
expressing the electric field through a chain rule in terms of derivatives
$\d/\d u_i^a$.  Providing a careful treatment of this problem is one of
the main goals of this paper. To establish that such a redundancy can
be handled, a detailed treatment will be presented of an example where there
remain in the presence of a large set of vector potentials, an infinite set
of such zero modes. We shall for that purpose propose an
infinitesimal deformation of \defi\ which, as we shall verify explicitly,
resolves the difficulties with zero modes.  In the sections that
follow, we shall first write down all the relevant geometrical formulas in
the presence of the deformation. Then, as an example of the treatment
where this care is required, we will specialize to those gauge
field configurations for which the associated geometries are 3-spheres,
where we shall be able to give the explicit expression for the
electric field.  In the limit in which the deformation is
eliminated, we shall see that rather than an indefinition in the
electric energy, there will be a restriction on the possible wavefunctionals
describing states of the theory for such configurations.
Finally, we shall give the expression for the electric field of a
system of infinitely heavy point color sources immersed in these spherical
configurations, using the formalism for introducing sources also
presented in I..

In introducing a deformation to \defi , again we are careful to preserve
both reparametrization and gauge covariance, and we must verify that it
indeed removes any zero mode ambiguities. We choose
\eqn\defdef{\ee^{ijk}D_ju_k^a=p\ee^{ijk}\ee^{abc}u_j^bu_k^c\ ,}
where $p$ is a small parameter with dimensions of mass. It is possible
to find $A_i^a$ explicitly as a functional of $u_i^a$. With manipulations
similar to those found in I., one finds
\eqn\aofu{A_i^a[u]=pu_i^a+ {(\ee^{nmk}\pa_mu^b_k)(u^a_nu^b_i-\half
u^b_nu^a_i) \over \det u}\ .}

One can already glean from the above why this
eliminates zero mode problems: variations $\d u, \d A$ must satisfy
\eqn\var{\ee^{ijk}(\d^{ac}\pa_j+\ee^{abc}
(A_j^b-2pu_j^b))\d u_k^c =-\ee^{ijk}\ee^{abc}u_j^b\d A_k^c\ .}
To obtain $\d u_i^a$ in terms of $\d A_i^a$, the operator acting on
$\d u_i^a$ must be inverted. With $p=0$ this operator would be
$\ee^{ijk}D_j$ itself, whose zero modes are the very solutions to \defi ,
while for $p\neq 0$, on the other hand,
possible zero modes of the operator on the l.h.s. of \var\ are clearly
not solutions to \defdef . In fact, the claim we shall make is that the
operator in \var\ has {\it no} zero modes for small enough nonzero $p$
and can always be inverted, leading to an unambiguous definition of the
electric field in the $u$-variables. We now proceed to present the relevant
formulas in geometric variables.\medskip

\newsec{Gauge Invariant Geometric Variables}\bigskip

With the definition \defdef, and using the gauge Ricci identity
analogously to what was done in I., one can determine that the magnetic
field, when expressed in terms of geometric variables, is
\eqn\bofu{B^{ai}\ =\ \sqrt{g}\ (G^{ij}+p^2g^{ij})\ u_j^a\ .}
Here $G_{ij}$ is the Einstein tensor of the metric $g_{ij}$, $G_{ij}=
R_{ij}-\half g_{ij}R$, and throughout indices are raised and lowered
with the metric $g_{ij}$ and its inverse $g^{ij}$. By
$\sqrt{g}$ we mean $\det u$, which can in principle take on both
positive and negative values. The gauge Bianchi identity can be worked
out to be
\eqn\bianchi{D_i B^{ai}\ =\ \sqrt{g}\ (\nabla_i G^{ij})\ u_j^a\ =0\ ,}
so that it implies the geometric Bianchi identity and vice-versa. We
note that this is a nontrivial consistency check of our geometric
picture.

We now turn to Gauss' law and the electric field. If we define,
following I., the gauge invariant tensor $e^{ij}$ through
\eqn\eij{iE^{ai}={\delta\over \delta A_i^a}
\ \equiv\ \sqrt{g}\ u_j^a e^{ij}\ ,}
we can then verify that the Gauss law generator in geometric variables
becomes
\eqn\gl{i{\cal G}^a =D_i \left({\delta\over \delta A_i^a} \right)
= \sqrt{g}u_j^a(\tilde{\nabla}_i e^{ij})=\sqrt{g}u_j^a(\nabla_i e^{ij})
+pu^{ia}\ee_{ijk}e^{jk}\ ,}
where $\tilde{\nabla}_i$ is the geometric covariant derivative with the
torsion term introduced by $p\neq 0$, and $u^{ia}$ is the matrix inverse
of $u_i^a$ (or, what amounts to the same
thing, $u_j^a$ with the space index raised
by $g^{ij}$). Alternatively, it follows
directly from \defdef\ that if one makes a gauge variation of $A_i^a$,
\eqn\gvari{\d A_i^a =- D_i\d w^a\ ,} then
\eqn\dtwop{\ee^{ijk}D^{(2p)}_j(\d u^a_k-\ee^{abc}\d w^bu^c_k)=0\ ,}
where the operator $D^{(2p)}$ contains
in place of $A$ the ``potential" $A-2pu$.
Since $\ee D^{(2p)}$ has no zero modes it follows that, as expected, a gauge
variation of $A$ is equivalent to a gauge
variation of $u$ transforming as a vector
so ${\cal G}^a$ is represented in terms of $u$ by
\eqn\gaussu{i{\cal G}^a=\ee^{abc}u^b_i{\d\over\d u^c_i}\ .}

To calculate the electric field in the $u$-variables, we begin from the
definition \eij\ of $e^{ij}$:
\eqn\chaina{e^{ij}={1\over\sqrt{g}}u^{aj}{\d\over\d A_i^a}=
{1\over\sqrt{g}}u^{aj}{\d u_k^b\over\d A_i^a}\cdot {\d\over\d u_k^b}\ ,}
where the dot stands for a generalized contraction including an
integration over space. Variations in $u_i^a$ can be further separated into
variations of the six gauge invariant degrees of freedom $g_{ij}$ and of
three gauge degrees of freedom in the following simple way: first we split
the summation over the index $b$ by inserting the unit color matrix in
the form $\d^{bc}=u^b_mu^{cm}$. The resulting
quantity $u^{cm}{\d\over\d u_k^c}$can then be written as the sum of its
symmetric and antisymmetric pieces in $m$ and $k$. Finally, it is easy to
see that these correspond, respectively, to variations in the
metric $g_{ij}$ and gauge variations. The final expression we arrive at is:
\eqn\chainc{e^{ij}(x)=\int\dy{1\over\sqrt{g}(x)}\left( u^{aj}(x)
{\d u_k^b(y)\over\d A_i^a(x)}u_m^b(y)\right)\left(2{\d\over\d
g_{km}(y)}+{i\over2}{\ee^{km\ell}\over \sqrt{g}(y)}u_\ell^a(y) {\cal
G}^a(y)\right)\ ,}
where $i{\cal G}^a=\ee^{abc}u^b_i\d/\d u^c_i$ is again the Gauss law
generator, now in terms of the $u$-variables, and we make
explicit the space integration.

We now need to write the Jacobian matrix $\d u/\d A$ in geometric form.
To do so, we start by considering the following eigenvalue problem:
\eqn\eigen{\ee^{ijk}(\d^{ac}\pa_j+\ee^{abc}(A_j^b-pu_j^b)){w_A}_k^c =
\sqrt{g}\l_A {w_A}^{ia}\ ,}
where again indices are raised with the inverse metric $g^{ij}=u^{ia}
u^{ja}$. We note that, by definition, one solution to the above with
$\l_A=0$ is $u_i^a$ itself. In the notation we are using, $A$ is an
index that labels all these eigenfunctions {\it except}
the particular one given by $u_i^a$. The operator
above is real and symmetric, and we assume $\{ u_i^a,{w_A}_i^a\}$ forms
a complete orthonormal spectrum of real eigenfunctions for it. By
orthonormality we mean
\eqn\ortho{\eqalign{\int\dx\sqrt{g}g^{ij}u_i^a {w_A}_j^a&=0\cr
\int\dx\sqrt{g}g^{ij} {w_A}_i^a {w_B}_j^a&=3V\d_{AB}\cr
\int\dx\sqrt{g}g^{ij}u_i^a u_j^a&=3\int\dx\sqrt{g}=3V\ ,}}
where $V$ is the volume of the space described by $g_{ij}$, and
$\d_{AB}$ is a Kronecker or Dirac delta depending on whether the
spectrum is discrete or continuous. Because we will eventually
concentrate on spherical geometries, we are only considering here
configurations of finite volume (that is the ``dynamical" volume $V$,
and not the volume of space, which is infinite). The generalization
to infinite $V$ should not entail further conceptual difficulty.

If we now expand a generic variation $\d u_i^a$ in terms of this
complete set,
\eqn\expansion{\d u_i^a=\eta u_i^a+\sum_A \eta_A{w_A}_i^a\ ,}
substitute this in \var\ and dot it on the left with the same complete
set (``dot" meaning an inner product with the measure
$\sqrt{g}g^{ij}$), we easily get the following relations:
\eqn\coeffs{\eqalign{3Vp\eta &=\int\dx\sqrt{g}u^{ia}\d A_i^a\cr
3V\sum_BI_{AB}\eta_B &=-\int\dx\ee^{ijk}\ee^{abc}u_i^{a}
{w_A}_j^b\d A_k^c\ ,}}
where
\eqn\IAB{I_{AB}=\l_A\d_{AB}-{p\over3V}\int\dx\ee^{ijk}\ee^{abc}u_i^{a}
{w_A}_j^b {w_B}_k^c\ .}
The origin of the zero mode problems alluded to above and their resolution
through the $p\neq0$ deformation of \defi\ now become manifest: for $p=0$,
the first of eqs. \coeffs\ actually represents a constraint on the
variations $\d A_i^a$ for which a $\d u_i^a$ can be found. This constraint
is a direct consequence of the fact that \aofu\ is homogeneous in
$u_i^a$ for $p=0$. For $p\neq0$, this homogeneity is clearly broken,
and there is no longer a constraint. Furthermore, essentially the same
happens with the second set of equations in \coeffs: for $p=0$ further
constraints on $\d A_i^a$ follow for each mode for which $\l_A=0$. Again,
these are eliminated by taking $p\neq0$.

{}From here, it is straightforward to write the Jacobian matrix in
explicit form:
\eqn\jacobian{{\d u_i^a(x)\over\d A_j^b(y)}={1\over3Vp}\sqrt{g}(y)
u^{jb}(y)u_i^a(x)-{1\over3V}\sum_{AB}I_{AB}^{-1}\ee^{jmn}\ee^{bcd}
u_m^c(y) {w_B}_n^d(y){w_A}_i^a(x)\ .}

This can be expressed in an entirely geometric form by using the important
fact that if ${w_A}_i^a$ is a mode of \eigen, then the geometric modes
${z_A}_i^{~j}$ defined through
\eqn\zmodes{{w_A}_i^a={z_A}_i^{~j}u^a_j}
can be seen to be eigenmodes of the geometric curl operator $\ee^{ijk}
\nabla_j$ with the same eigenvalues $\l_A$:
\eqn\curl{\ee^{ijk}\nabla_j{z_A}_k^{~m}=\sqrt{g}\l_A {z_A}^{im}\ .}
This leads to the fully geometric form we were seeking for the operator
appearing in \chainc:
\eqn\geomjacob{\left( u^{aj}(x){\d u_k^b(y)\over\d A_i^a(x)} u_m^b(y)
\right)\! =\!\sqrt{g}(x)\!\left[{1\over 3Vp}g^{ij}(x)g_{km}(y)+
{\cal H}_{~~km}^{ji}(x,y)-g^{ij}(x){\cal H}_{s~km}^{~s}(x,y)
\right]\ ,}
where the Green's function ${\cal H}_{ijmn}(x,y)$ is defined to be:
\eqn\green{ {\cal H}_{ijmn}(x,y)\equiv {1\over 3V}
\sum_{AB}{z_A}_{ij}(x) I_{AB}^{-1}{z_B}_{mn}(y)\ .}
Assembling all these results leads to the following electric geometric
tensor acting on functionals $\Ps $:
\eqn\eijfinal{\eqalign{e^{ij}(x)\Ps =\int\dy
&\left[{1\over 3Vp}g^{ij}(x)g_{mn}(y)+
\left( {\cal H}_{~~mn}^{ji}(x,y)-g^{ij}
(x){\cal H}_{s~mn}^{~s}(x,y) \right)\right]\cdot\cr &\phantom{xxxxxxx}
\left( 2 {\d\Ps\over\d g_{mn}(y)}+{i\over2}{\ee^{mn\ell}\over
\sqrt{g}(y)}u_\ell^a(y) {\cal G}^a(y)\Ps\right)\ .}}
{}From this expression, one may already observe that, independent of the
geometry, there is always at least one divergence in the electric
energy as $p\to0$. We eliminate it by requiring that gauge invariant
functionals $\Ps [u]$ be invariant under global rescalings of the
metric, {\it i.e.},
\eqn\resc{\int\dy u^a_i(y) {\d\Ps\over\d u^a_i(y)}=0\ .}

The Green's function ${\cal H}_{ijmn}(x,y)$ may also have divergences
in the limit $p\to0$, which again have to be eliminated. Generally
speaking, the higher the degree of symmetry of a certain geometry
(which is determined by its Killing vectors), the larger the number
of zero modes of the curl operator and, due to \IAB, the larger the
number of divergent terms in ${\cal H}_{ijmn}$ as $p\to0$.
In the following section we will work out and analyze the electric
field for those compact geometries with the maximum number of
Killing vectors, namely, spheres. Because they are
maximally symmetric spaces, for spheres it is possible to find
the spectrum of $\ee\nabla$ explicitly
without too much difficulty, and therefore an explicit
expression for ${\cal H}_{ijmn}$ as well.

A final note on renormalization of divergences is in order here. It has
been  usual in the past literature on the subject to consider as the electric
energy density expectation on a state $\Ps$ the expression
\eqn\elena{<\Ps |(E^{ai}(x))^2|\Ps >=-\int [{\cal D}\! A]\ \Ps {\d^2\over
\d A_i^{a}(x)^2}\Ps\ ,}
with $[{\cal D}\! A]$ an appropriately defined integration measure. However,
another way to define the electric energy expectation is (cf. I.)
\eqn\elenb{<\Ps |(E^{ai}(x))^2|\Ps >=\int [{\cal D}\! A]\ {\d\Ps\over \d
A_i^{a}(x)}{\d\Ps\over \d A_i^{a}(x)}\ .}
The former expression is inherently divergent due to the coincident points
in the double functional derivative, and one must go to some lengths to
properly define the operator. The latter expression, on the other hand,
is easier to define and can be seen as an alternative prescription for
the electric energy density. In our work we always use this second form.
Of course, for systems with a finite number of degrees of freedom the two
expressions are equivalent. Here, they differ formally by a total
functional derivative.\medskip

\newsec{Spherical Configurations}\bigskip

In order to get a clear picture of our restriction to spherical
configurations, the first questions we will address are $a)$ to
what $A_i^a(x)$ configurations do spherical geometries correspond,
and $b)$ how much of the entire space of $A_i^a(x)$ do these geometries cover.

The direct way of answering the first question is of course to
take a configuration $u_i^a(x)$ describing a 3-sphere and
substitute it in \aofu\ to find $A_i^a(x)$. We will do so below
for a particular metric on $S_3$. There is, however,
a more indirect but extremely economical way,
based on the following reasoning: the geometry of a
sphere is that of an Einstein space, for which $G_{ij}\propto g_{ij}$;
this implies, by \bofu, that the magnetic field for such geometries
must be proportional to the matrix inverse of the 3-bein,
$B^{ai}\propto u^{ai}$. If we now consider the standard expression
\bofa\ for $B^{ai}$ as a function of $A_i^a$, and the fact that pure
gauge configurations, say $\bar{A}_i^a$, have vanishing magnetic
field, it follows immediately that for configurations which are global
scalings of a pure gauge, say  ${A}_i^a(x)=k\bar{A}_i^a(x),
k\neq1$, the magnetic field will turn out to be proportional to
$B^{ai}\propto\ee^{ijk}\ee^{abc}{A}_j^b{A}_k^c$. But
this is proportional to the matrix inverse of ${A}_i^a$ and thus,
for spherical geometries, $A_i^a(x)$ must be
proportional to  $u_i^a(x)$. Closer scrutiny of this argument shows
that it indeed holds and furnishes all the proportionality constants
missing above. We now list a series of results that derive
from the above reasoning. In what follows, we take $\bar{A}_i^a(x)$
to be a pure gauge configuration, and $\a$ a real number $\neq1$.\medskip

\item {\it i)} $A_i^a(x)={1\over2(1-\a )}\bar{A}_i^a(x)\Longleftrightarrow
u_i^a(x)={\a \over p}A_i^a(x)={\a\over 2p(1-\a )}\bar{A}_i^a(x)$.
\medskip

\item {\it ii)} $A_i^a(x)={1\over2(1-\a )}\bar{A}_i^a(x)\Longrightarrow
u_i^a(x)$ is an Einstein space, with $G_{ij}(x)=-{R\over6}g_{ij}(x)
=-p^2 {(1-\a )^2\over \a^2}g_{ij}(x)$.\medskip

\item {\it iii)} $u_i^a(x)$ is an Einstein space, with $G_{ij}(x)=-
{R\over6}g_{ij}(x)=-p^2 {(1-\a )^2\over \a^2} g_{ij}(x)
\Longrightarrow {\cal A}_i^a(x)\equiv A_i^a(x)-{p\over \a}u_i^a(x)$ is
pure gauge.\medskip

\item {\it iv)} ${\cal A}_i^a(x)\equiv A_i^a(x)-{p\over \a}u_i^a(x)$ is
pure gauge $\Longrightarrow (A^g)_i^a(x)={p\over \a}(u^g)_i^a(x)$ for
some gauge transform $(A^g,u^g)$ of $(A,u)$.\medskip

Thus, up to gauge transformations, the circle of implications
above flows freely in both directions. Moreover, we can now
also answer the second question as well: the space of all spherical
geometries corresponds to the space of all vector potentials
that are rescalings of all possible pure gauges. More concretely,
all possible pure gauges in $SU(2)$ are spanned by three real
functions $\xi^a(x)$, and rescalings are spanned by one real
number.\footnote{$^\dagger$}{Incidentally, the special case of $\a =1$
can be treated separately, and is seen to correspond to a flat
geometry where, in an appropriate gauge, $u_i^a(x)=\d_i^a$ and
the vector potential is  $A_i^a(x)=p\d_i^a$.}
This is indeed expected and consistent with the geometric
picture, since all possible 3-sphere metrics are spanned by
one real number (the inverse radius of the sphere, in units of $p$)
and three real functions $y^i(x)$ (coordinate reparametrizations
of a reference metric). This should be contrasted with six real,
local functions, which parametrize the physical, gauge invariant,
Hilbert space of the theory, so that, roughly speaking, 3-spheres
span half the dimensions of this space. It is also possible to
study the case of noncompact maximally symmetric spaces, {\it i.e.},
3-hyperboloids, although we will not consider these geometries
here. The vector potential can easily be found by substituting
the appropriate bein $u_i^a(x)$ in \aofu. The procedure described
above for spheres can be extended to hyperboloids
by taking $\a$ complex, which would lead to complex potentials.
Insofar as the symmetries are concerned, complex potentials do not
spoil any of the reasoning above; however, in order to have real
vector potentials in the end would require a complex gauge
transformation in {\it iv)} above. Altogether we know this is possible
since it is not difficult to obtain a real vector potential for this case.

To give a concrete example, we can consider the projective metric on
$S_3$ used below (cf. (3.16) and below for the coordinate conventions).
It is not difficult to find that in the limit of interest to us, $p\to0$, the
associated gauge field configuration is
\eqn\wymon{A_i^a(x)=-2\ {\ee_{iaj} x^j\over a^{-2}+|x|^2} \ .}
Here, $a^{-1}$ is the radius of the sphere. Two features of these
configurations are worthwhile noting: first, if we take $a^2$ negative,
we simply get the hyperbolic configurations mentioned above; unlike spherical
configurations, they have singularities at finite $x$. Secondly, these
configurations correspond precisely to the spatial components of the instanton
of Belavin {\it et al.}, with $a^{-1}$ the ``size" of the instanton \bps.
Further, a closely related type of configuration for which such a geometrical
picture is also readily available correspond to Wu-Yang monopoles \fk\ (which
are in turn related to merons \cdg). They are gotten by
taking $a^{-1}\to 0$ in the instanton configuration above, and multiplying
it by a global factor $\half$. The associated gauge invariant metric
variable is $g_{ij}=\r^2/|x|^2 \d_{ij}$, with $\r$ a constant parameter. This
metric again describes a space of constant curvature, but closer inspection
of its curvature invariants reveals that
it actually corresponds to the space $S_2\times\IR$. It has been argued that
(coherent states of) these magnetic monopoles are in fact one of the key
ingredients underlying the color confinement mechanism in QCD \man.
We will not pursue such a geometry further in this paper, but we wanted
nonetheless to illustrate the point that our geometric setting fits very
nicely with specific gauge field configurations that are deemed to be
important for the dynamics of Yang-Mills theory.

It is also curious to note that for spherical configurations, and again
in the limit $p\to0$, two of our fundamental equations are identical to
the two equations defining the classical phase space of $d=3$ gravity
in the presence of a cosmological constant. For $u_i^a$ seen as a dreibein
and $\o_i^{ab}\equiv\ee^{abc}A_i^c$ seen as a spin connection, Eq. \defi\
states that the connection is torsion free, and \bofu\ restricted to
spheres states that the curvature built out of the spin connection is
proportional to the inverse dreibein. These are just the equations of
motion of $d=3$ Einstein-Hilbert gravity with a cosmological constant
(related to $p^2$ in \bofu) or, equivalently, of an $SO(4)$ Chern-Simons
action with the gauge field being given by a combination of both $u_i^a$
and $\o_i^{ab}$ \witten. This analogy, of course, does not go any further
since it makes no mention of the electric field or of the specific form of the
Hamiltonian.

We now present our results on the spectrum of the curl operator \mrst.
The most straightforward way we found of calculating this spectrum
was by writing {\it ans\"atze} for the modes ${z_A}_{ij}$ based on
the scalar, vector and 2-tensor eigenmodes of the Laplacian on the
3-sphere and, after exhausting all possibilities for covariant
{\it ans\"atze}, verifying that the resulting modes ${z_A}_{ij}$
satisfy a completeness relation. The eigenmodes we find are:\medskip

\noindent
{\it 1. Exact zero modes.}
\eqn\zerom{{z_A}_{ij}=(Z_N)_{ij}={1\over a^2\sqrt{(\o^2_N-3)(\o^2_N
-1)}} (\nabla_i\nabla_j+a^2g_{ij})y_N\ , N=0,2,3,\ldots\ .}
Here and in what follows, $\o^2_N=N(N+2)$ is the spectrum of the
Laplacian acting on scalars on the 3-sphere, $y_N$ are the associated
eigenmodes (the hyperspherical harmonics) normalized to $3V$, and
$a$ is the inverse radius of the sphere (with scalar curvature
$R/6=a^2$), so that
\eqn\scalarlap{(\nabla^2+a^2\o_N^2)y_N=0\ .}
The index $A$ represents the three quantum numbers $(N,\ell,m)$
labelling these modes. Because $\ell$ and $m$ do not affect the
spectrum, but only its degeneracy, they are of secondary importance
here, and will be omitted and understood whenever possible.
The normalization is such that geometrical modes satisfy the normalization
conditions following from \ortho. For all these modes, $\l_A=0$. We
also note that $N=1$ is missing: it vanishes identically due to the
structure of the operator $(\nabla\nabla +a^2g)$. Moreover,
the $N=0$ term will also be eliminated from the calculation of
the Green's function in \green\ because it corresponds to the metric
zero mode $g_{ij}$, which is to be treated separately (cf.
discussion above \ortho ).\medskip

\noindent {\it 2. Scalar-based modes.}
These are nonzero modes based on the spectrum of the Laplacian acting
on scalars:
\eqn\scalarm{{z_A}_{ij}=(S_{N\a})_{ij}={1\over 2a^2\o_N
\sqrt{\o^2_N-1}} \left(\nabla_i\nabla_j+a^2\o^2_Ng_{ij}-\a a
\sqrt{\o^2_N-1}{\ee_{ijk}\over\sqrt{g}}\nabla^k\right)y_N\ .}
Here, $\a=\pm$, and the associated eigenvalues are $\l_A=\l_{N\a}=
\a a\sqrt{\o^2_N-1}$, $N=1,2,3,\ldots$.\medskip

\noindent
{\it 3. Vector-based modes.} These modes are based on the spectrum
of the Laplacian acting on vectors,
\eqn\vectorlap{(\nabla^2+a^2(\o_N^2-1))(v_N^{\a})_i=0\ ,}
with $\nabla^i(v_N^{\a})_i=0$. Such vectors are also eigenvectors of the
curl operator,
\eqn\vectorcurl{ {\ee_{ijk}\over\sqrt{g}}
\nabla^j(v_N^{\a})^k =\a a\sqrt{\o^2_N-1}(v_N^{\a})_i\ ,}
with $\a =\pm$. The vector-based modes are then given by
\eqn\vectorm{{z_A}_{ij}=(V_N^{\a\b})_{ij}={1\over 2}
\left[ A_N^{\a\b}\left(\nabla_i(v_N^{\a})_j+\nabla_j(v_N^{\a})_i
\right)-{\ee_{ijk}\over\sqrt{g}}(v_N^{\a})^k\right] \ ,}
with eigenvalues
\eqn\vectoreigen{\l_N^{\a\b}={\a a\over2}\left(\sqrt{\o^2_N+1}
+\b\sqrt{\o^2_N-3}\right)={\a a\over2}\left( N+1+\b\sqrt{(N+3)(N-1)}
\right)\ ,}
where $\b=\pm$, $A_N^{\a\b}=\a\b /(a\sqrt{\o^2_N-3})$,
and $N=2,3,4,\ldots$ (the $N=1$ modes built in this way vanish
identically, similarly to what happens in the exact zero mode
case).\medskip

\noindent
{\it 4. Tensor-based modes.} These modes are based on the
spectrum of the Laplacian acting on tensors,
\eqn\tensorlap{(\nabla^2+a^2(\o_N^2-2))(T_{N\a})_{ij}=0\ ,}
with $\nabla^i(T_{N\a})_{ij}=0$, $(T_{N\a})_i^i=0$ and $(T_{N\a})_{ij}$
symmetric. They are given by the $(T_{N\a})_{ij}$ themselves,
with $\a=\pm, N=2,3,4\ldots$ and eigenvalues $\l_{N\a}=\a a\sqrt{\o^2_N+1}
=\a a(N+1)$. Again, the putative $N=1$ mode vanishes identically, and the
spectrum starts from $N=2$.\medskip

It is now simple (but lengthy!) to calculate the matrix $I_{AB}$,
through \IAB\ (with the gauge modes $w$ referred to geometric modes
$z$ through \zmodes ). If we organize the matrix into five sectors,
$T$ (for tensor-based modes), $V$ (for vector-based modes), $S+$ (for
scalar-based modes with $\a=+$), $S-$ (for scalar-based modes with
$\a=-$), and $Z$ (for zero modes), a simple structure emerges: the
matrix is diagonal in the $T$ and $V$ sectors, and the only non-diagonal
couplings appear between the $Z$, $S$+ and $S-$ sectors. Furthermore, its
subblocks are diagonal in each and all of its nonvanishing sectors
(e.g., $TT$, $VV$, $ZS+$, $ZS-$, $ZZ$, etc.). This allows for an explicit
inversion, even though the matrix is infinite. The nonvanishing
entries of $I_{AB}$ in the different sectors are:
\eqn\ttt{I^{(TT)}_{N\a ,M\b}=(\a a\sqrt{\o^2_N+1}+p)\d_{MN}\d_{\a\b}}
\eqn\vv{I^{(VV)}_{N\a\b ,M\a '\b '}={\a a\over2}\left(\sqrt{\o^2_N+1}
+\b\sqrt{\o^2_N-3}\right)\d_{MN}\d_{\a\a '}\d_{\b\b '}\ ,}
(where $\d_{MN}$ includes orthogonality in $\ell ,m$ as well) and
\eqn\matrixi{I={p\over (\o^2_N-1)} \pmatrix{2&\o_N\sqrt{\o^2_N-3}
&\o_N\sqrt{\o^2_N-3}\cr \o_N\sqrt{\o^2_N-3}&{a\over p}
(\o^2_N-1)^{3/2}-\o_N^2&-1\cr \o_N\sqrt{\o^2_N-3}&-1&
-{a\over p}(\o^2_N-1)^{3/2}-\o_N^2\cr}\ ,}
in the $Z,S+,S-$ sectors, where the first, second and third row
(or column) refers to, respectively, $Z,S+$ and $S-$ sectors.
Each entry represents an infinite diagonal matrix and because
of this, inversion can be accomplished by simply inverting the
$3\times3$ matrix. This inverse is
\eqn\matrixiinv{\eqalign{I^{-1}&={a^2\over 2(p^2-a^2)\xi_N^2}
\cdot\cr &\pmatrix{p(\o^2_N+1)-{\xi_N^4\over a^2p}&
(p+\xi_N)\o_N\sqrt{\o^2_N-3} &(p-\xi_N)\o_N\sqrt{\o^2_N-3}\cr
(p+\xi_N)\o_N\sqrt{\o^2_N-3}& -2\xi_N-p\o_N^2&
p(\o^2_N-2)\cr (p-\xi_N)\o_N\sqrt{\o^2_N-3}&
p(\o^2_N-2)&2\xi_N-p\o_N^2\cr}\ ,}}
with $\xi_N=a\sqrt{\o^2_N-1}$.

These results can finally be substituted in \green\ in order
to calculate ${\cal H}_{ijmn}(x,y)$. The $TT$ and $VV$
contributions can easily be gleaned from \ttt\ and \vv, while the
contribution from the $Z,S+,S-$ sectors is quite lengthy. In
fact, a series of simplifications take place, and the final
result is:
\eqn\greenfinal{\eqalign{{\cal H}_{ijmn}(x,y)&=
{1\over 2a^2p} (\nabla_i\nabla_j+a^2g_{ij})^x
(\nabla_m\nabla_n+a^2g_{mn})^y G_3(x,y)+\cr
-{p\over 2a^2(p^2-a^2)}&\left[(\nabla_i\nabla_j)^x
(\nabla_m\nabla_n)^y +\left({a\ee_{ijk}\nabla^k\over
\sqrt{g}}\right)^x\left({a\ee_{mn\ell}\nabla^\ell\over
\sqrt{g}}\right)^y\right]G_0(x,y)\cr
-{a\over 2a^2(p^2-a^2)}&\left[(\nabla_i\nabla_j)^x
\left({a\ee_{mn\ell}\nabla^\ell\over
\sqrt{g}}\right)^y+\left({a\ee_{ijk}\nabla^k\over
\sqrt{g}}\right)^x(\nabla_m\nabla_n)^y\right]G_0(x,y)+
\ldots\ .}}
The dots represent the $TT$ and $VV$ contributions,
and $G_0$ and $G_3$ are the Green's functions for the operators $\nabla^2$
and $(\nabla^2+3a^2)$, respectively, acting on scalars on the sphere:
\eqn\gc{\eqalign{G_0(x,y)\equiv &{1\over3V}\sum_{N=1}^\infty {y_N(x)y_N(y)
\over a^2\o_N^2}\cr G_3(x,y)\equiv &{1\over3V}\sum_{N=2}^\infty {y_N(x)y_N(y)
\over a^2(\o_N^2-3)}\ .}}
Since $(\nabla^2+3a^2)$ has zero modes given by $y_{1\ell m}(x)$, it does
not strictly speaking have a Green's function;
what we mean by the above is of course the
Green's function on the subspace of functions on
$S_3$ that is orthogonal to this zero mode.
In fact, both Green's functions above are also
lacking the trivial $y_0={\rm const.}$
mode in their spectral sum. This will be
reflected in the differential equations they satisfy.

It is in fact possible, with some effort, to
find closed expressions for these propagators.
We present them in what follows and briefly
describe how they are gotten since these are
useful in the calculations envisaged in Sec.
4. We use the standard projective metric on $S_3$:
\eqn\metrics{g_{ij}(x_1,x_2,x_3)={4\over(1+a^2|x|^2)^2}\d_{ij}\ ,}
where $x_1,x_2,x_3$ are projective coordinates, with range $-\infty$ to
$\infty$, $|x|^2=x_1^2+x_2^2+x_3^2$, and $a$ is the inverse radius of the
sphere. From this, the Laplacian acting
on scalars on $S_3$ can be built, leading
to the following differential equations for $G_0$ and $G_3$:
\eqn\laplg{\eqalign{\nabla^2 G_0(x,y)=&-{\d (x-y)\over
\sqrt{g}(x)}+{1\over3V}y_0(x)y_0(y)=-
{\d (x-y)\over\sqrt{g}(x)}+{1\over V}
\cr (\nabla^2+3a^2)G_3(x,y)=&-{\d (x-y)\over \sqrt{g}(x)}+{1\over3V}
\sum_{N=0}^{1}\sum_{\ell=0} ^{N}\sum_{m=-\ell}^{\ell}y_{N\ell m}(x)
y_{N\ell m}(y)\ ,}}
where $V=2\p^2/a^3$ is the volume of the sphere.
The extra terms on the r.h.s.~represent
the lack of completeness of these Green's
functions, as alluded to above. We will
not go into the lengthy details of calculation,
but rather just present the final result for $G_0$ and $G_3$:
\eqn\propfinalz{G_0(x,y)={a\over8\p}\left( {\sqrt{1-\eta^2}\over\eta}
-{\eta\over\sqrt{1-\eta^2}}\right)(1-{2\over\p}\sin^{-1}
\eta)-{a\over8\p^2}}\eqn\propfinalth{G_3(x,y)={a\over16\p}
\left({1\over\eta\sqrt{1-\eta^2}}-8\eta\sqrt{1-\eta^2}\right)(1-{2\over\p}
(\sin^{-1}\eta -\cos^{-1}\eta))+{a\over24\p^3}(6\eta^2+1)\, }
where
\eqn\chord{\eta\equiv{ad\over2}={a|x-y|\over\sqrt{1+a^2|x-y|^2}}}
is one half the chordal distance $d$
between the points $x$ and $y$ in units of $a^{-1}$.

We are in fact interested in the $p\to0$ limit of \greenfinal,
and there are a few important features to note regarding
this limit. Firstly, in the $TT$ and $VV$ sectors, the
$p\to0$ limit is perfectly smooth; in particular, we would
have obtained the same result had we taken $p=0$ from the
beginning. Secondly, we find in the $ZZ$ sector another
$1/p$ divergence exactly like the one associated with
global scalings (cf. \eijfinal\ and \resc).
This divergence appears in the first term in
\greenfinal, and it will likewise entail a
constraint on finite energy physical wavefunctionals.
What this constraint is can be seen from the term
$g^{ij}{\cal H}_{s~~mn}^{~~s}$ in \eijfinal: the
trace in the first two indices leads to the
operator $(\nabla^2+3a^2)$ acting on $G_3$,
and this leads to three types of terms, as
can be seen from \laplg. The first term is a
$\d (x-y)$, and vanishing of this term leads to the constraint
\eqn\extracond{\left. (\nabla_m\nabla_n+a^2g_{mn})^y
{\d\Ps\over\d g_{mn}(y)}\right|_{g_{ij}={\rm sphere}}=0}
in order not to have a divergence in the
limit $p\to 0$. The second term contains
$y_0(x)y_0(y)={\rm const.}$, and the second
operator $(\nabla \nabla +a^2g)$ acting on
it kills it because of the constraint
\resc. The third type of term contains
$y_{1\ell m}(x)y_{1\ell m}(y)$, and these
vanish automatically under the action of
the second operator $(\nabla \nabla +a^2g)$.
Thus, only the first term leads to a constraint, \extracond, on physical
wavefunctions.This constraint must be satisfied by
physical wavefunctionals in order for their energy to be finite in the limit
$p\to0$.This is again a result we would obtain
directly in the $p=0$ case from a similar requirement of
finiteness of the electric energy. Finally, we note that
the last term in \greenfinal\ above is finite and
nonvanishing in the  $p\to0$ limit. This result is different
from what one would obtain by treating the $p=0$ case
directly. The correct result is the one presented here, since
it properly takes into account the mixing of the zero and scalar modes.

In the $p\to0$ limit and with the above finiteness constraints
in place, the Green's function ${\cal H}_{ijmn}$ is
\eqn\greenfull{\eqalign{&{\cal H}_{ijmn}(x,y)={1\over3V}\sum_N \left\{
{1\over a\sqrt{\o_N^2-1}} \left[ (T_{N+})_{ij}(x)(T_{N+})_{mn}(y)-
(T_{N-})_{ij}(x)(T_{N-})_{mn}(y)\right]\right.\cr
&+{2\over a(\sqrt{\o_N^2+1}+\sqrt{\o_N^2-3})} \left[
(V_{N++})_{ij}(x)(V_{N++})_{mn}(y)-
(V_{N-+})_{ij}(x)(V_{N-+})_{mn}(y)\right]\cr
&\left. +{2\over a(\sqrt{\o_N^2+1}-\sqrt{\o_N^2-3})}
\left[ (V_{N+-})_{ij}(x)(V_{N+-})_{mn}(y)-
(V_{N--})_{ij}(x)(V_{N--})_{mn}(y)\right]\right\}\cr
&+{1\over 2a^2}\left[ (\nabla_i\nabla_j)^x
\left({\ee_{mn\ell}\nabla^\ell\over
\sqrt{g}}\right)^y+\left({\ee_{ijk}\nabla^k\over
\sqrt{g}}\right)^x(\nabla_m\nabla_n)^y\right]G_0(x,y)\ ,}}
while its trace reduces to a single term
\eqn\greentrace{{\cal H}_{s~~mn}^{~~s}(x,y)=-
{1\over 2a^2}\left({\ee_{mn\ell}\nabla^\ell\over
\sqrt{g}}\right)^y {\d (x-y)\over \sqrt{g}(x)}\ .}
{}From now on we assume this limit unless
stated otherwise. We now have an explicit expression for both electric and
magnetic energy densities. Although the electric energy is still in a
rather unwieldy form, it is possible already to make an important
observation regarding the vacuum state of the theory: spherical
configurations introduce a scale into the problem, as any other explicit
configuration would. Such a scale must be dynamically determined and,
although we will not perform such a calculation here, we can already
observe that this scale $a$ enters the magnetic energy with a positive
power and the electric energy in negative
powers. This will cause the ground state
wave functional to fall rapidly for large
amplitude magnetic densities which vary
slowly in space. At the same time, it will
also become small for low amplitude magnetic
energy densities with slow spatial variation.
The correspondingly reduced fluctuations in
the magnetic energy density will presumably
fill the role of what is meant by a magnetic
``condensate". If this is correct then one should be able to get at least a
semi-quantitative estimate of the long range
color electric fields produced by static sources
in such surroundings. To do this we suggest looking at the
electric field by dropping terms which should mainly be associated with short
scales.

It would also be possible to perform manipulations in the vector and tensor
sectors similarly to what has been done for the scalar and zero sectors,
in order to simplify their respective contributions to ${\cal H}_{ijmn}$.
However, as we shall argue, it is these latter sectors ({\it i.e.},
the $\nabla\nabla\nabla G_0$ term in \greenfull ) which will give
rise to the main contribution to the potential between color sources,
and therefore one may in fact drop the $TT$ and
$VV$ terms in a first approximation. To identify
particular terms in ${\cal H}_{ijmn}$ that
lead to large electric energy densities, we
must look both for eigenvalues $\l_A$
such that $1/\l_A$ becomes large, and for modes which do not oscillate
much, since highly oscillating modes cannot contribute to long distance
effects. The third term in \greenfull\ (the last $VV$ term) does have
asymptotically large inverse eigenvalues as
$N\to\infty$; however, these are also associated to highly oscillatory modes.
On the other hand, all theslowly oscillating $T$ and $V$
modes do not have inverse eigenvalues that
become asymptotically large. The only term satisfying both conditions we
are seeking is the last one, associated to the scalar and zero sectors, and
therefore it is reasonable to keep only this term as a leading approximation.
Naturally, our formalism automatically guarantees, as announced in Sec.
$1$, that this represents a {\it gauge invariant} approximation to the
dynamics. \medskip

\newsec{Static Point Color Sources}\bigskip

We now consider the energy density of infinitely heavy
point color sources immersed in the Yang-Mills configurations
associated to spherical geometries. The formalism for
introducing point color sources has been developed in I..

Introducing color sources at isolated points in space entails
a local modification of Gauss' law only at these points. Then,
rather than introducing an additional set of variables at
every point in space, one can accomodate these isolated
inhomogeneities in Gauss' law by simply considering wavefunctionals
that carry the appropriate representation for each source,
but that are still functionals of $u_i^a$ only. To be specific,
let us consider, for instance, the insertion of two sources
at points $x_1$ and $x_2$. The generalization
to more sources is entirely trivial, but we
consider here this specific case for clarity of presentation.
Then, wavefunctionals describing states of this system
should take the form
\eqn\funcform{\Ps_{\a\b}[u_i^a]\ ,}
where $\a$ is an index in some $SU(2)$ representation transforming
at point$x_1$ and $\b$ likewise, but transforming at $x_2$. The
modification in Gauss' law is
\eqn\modgauss{ {\cal G}^a(x)\rightarrow \bar{{\cal G}}^a(x)
_{\a\a '\b\b '}={\cal G}^a(x)\d_{\a\a '}\d_{\b\b '}+
\L^a_{\a\a '}\d_{\b\b '}\d (x-x_1)+
\d_{\a\a '}\L^a_{\b\b '}\d (x-x_2)\ ,}
where $\L^a$ are the appropriate $SU(2)$
generators. Again, to be specific, we consider the
two sources to be in the fundamental
representation, in which case the $\L^a$
are proportional to the Pauli matrices
$\s^a$. The statement of gauge invariance becomes
\eqn\gisource{\bar{{\cal G}}^a(x)_{\a\a '\b\b '}\Ps_{\a '\b '}[u_i^a]=0\ .}

At this point it may not be clear whether or how one can build a color
singlet wavefunctional, satisfying the {\it local} constraint \modgauss,
exactly at the locations $x_1$ and $x_2$ of the sources, since at each of
these points the total color has contributions coming only from the
combination of ``integer spin" variables in the adjoint representation
(the $u_i^a$), and a ``half-integer spin" variable ($\a$ or $\b$) coming
from the source, which is in the fundamental representation. This turns out
to be possible because there are sufficient variables $u_i^a$ in order to
build a half-integer spin representation of
$SU(2)$ at $x_1$ and at $x_2$ with these
variables alone, even though they are in the adjoint representation. The
way this is done is by realizing that $u_1^a,u_2^a$ and $u_3^a$ form
three vectors, and thus comprise nine degrees of freedom at the point $x_1$
or $x_2$.
While six of these degrees of freedom, $u_i^au_j^a$ ({\it i.e.}, the gauge
invariant ones), give three lengths and three angles with which to uniquely
define a tetrahedron, the three remaining degrees of freedom can be
used to define the Euler angles uniquely
fixing the orientation of the tetrahedron
in color space. We then make use of the fact that
it is possible to build half-integer spin representations of SU(2) with
three Euler angles. With the appropriate transformation of
variables from $u_i^a$ to Euler angles, the angular momentum operator becomes
precisely ${\cal G}^a$, and the appropriate eigenfunctions are the Wigner
${\cal D}^{(j)}$-functions, with $j=\half$ for angular momentum one-half.
Equations \modgauss,\gisource\ then express
the fact that  under the usual addition of
angular momentum in quantum mechanics, the
wavefunctional at $x_1$ and at $x_2$ is a
singlet, of total angular momentum zero,
built out of two spin one-half representations.
Because this procedure is to be done at the isolated points
$x_1$ and $x_2$, that is, because of the delta functions $\d (x-x_1)$ and
$\d (x-x_2)$, the wavefunctional, besides being a functional of $u_i^a(x)$
everywhere, must now also be a regular function of the variables
$u_i^a(x_1)$ and $u_i^a(x_2)$:
\eqn\psisource{\Ps =\Ps_{\a\b}[u_i^a;u_i^a(x_1),u_i^a(x_2)]\ ,}
so that the functional differentiation in ${\cal G}^a$ at those points
becomes a regular derivative, and automatically incorporates the two
delta functions.

With the introduction of sources, the change in magnetic energy
is due not to any modification in \bofu\ but rather to the
constraints on $\Ps$ engendered by \gisource. For the electric
energy, on the other hand, there are modifications coming from
the Gauss law term in \eijfinal, which would otherwise be absent
for gauge invariant functionals. Neglecting the $TT$ and $VV$ contributions
to ${\cal H}_{ijmn}$ in \greenfull,
the contribution of the sources to the geometric electric tensor is
\eqn\eijsource{\eqalign{a^2e^{\rm source}_{ij}(x)
\Ps &=\int\dy \left\{\left[ (\nabla_i\nabla_j)^x
\left({\ee_{mn\ell}\nabla^\ell\over\sqrt{g}}\right)^y-\left({\ee_{ijk}
\nabla^k\over\sqrt{g}}\right)^x(\nabla_m\nabla_n)^y
\right]G_0(x,y)\right.\cr &\left.-g_{ij}(x)\left[\nabla_x^2\left(
{\ee_{mn\ell}\nabla^\ell \over\sqrt{g}}\right)^y\right]G_0(x,y)
\right\} \left( {\ee^{mnp}\over \sqrt{g}(y)}{\cal G}_p(y)\Ps\right)\ ,}}
which simplifies to
\eqn\eijsourceb{2a^2e^{\rm source}_{ij}(x)\Ps [g]
=[(\nabla_i\nabla_j)^x\nabla_k^{x_1}G_0(x,x_1)]
\L^k(x_1)\Ps\! +\! g_{ij}(x){1\over\sqrt{g}(x)}(\nabla_k^{x_1}\d (x-x_1))
\L^k(x_1)\Ps }
plus an identical contribution at $x_2$.
Here, $\L_k(x)\equiv -iu_k^a(x)\L^a$,
and we have ommited the $SU(2)$ indices. The action of
${\cal G}^a$ on $\Ps$ has been such as to satisfy \gisource.

One could now calculate an expression for the ``potential" associated with
static sources just as one evaluates the static Coulomb energy in the abelian
gauge theory. It is clear that as well as being more complicated, the
static potential is a function of the gauge field configuration and hence even
in an approximation of the Born-Oppenheimer type it must be
averaged with the ground state wavefunctional of the gauge field.
Here we have obtained an explicit expression only for those gauge field
configurations which are related to each other by $GL(3)$ transformations of
equal curvature geometries.
We shall postpone a more complete discussion, and an explicit
evaluation of the ``potential" associated with the electric field
in \eijsourceb\ for a later publication.

\medskip

\newsec{Conclusions}\bigskip

In this paper we have pursued further the formalism developed in I.,
where a set of local gauge invariant variables were introduced to
describe the physical Hilbert space of Yang-Mills theory. We have
chosen to do this in a Hamiltonian, fixed-time formalism because
there one can identify the subset of the full gauge group that truly
acts as a quantum mechanical symmetry of the theory, and one can implement
it in a manifest and exact way. We have furthermore showed that the
present treatment allows for approximations to the dynamics that
do not spoil this exact gauge symmetry.

As a first step towards concrete calculations, we have worked out in
detail the expression for the electric and magnetic energies for those gauge
field configurations corresponding to spherical geometries. This has
furnished indications of the mechanism through which a dynamically
determined scale enters the theory and leads to a nonvanishing
magnetic energy density of the vacuum and a mass gap. We have also indicated
explicitly how such a geometry is related to instanton configurations,
and how magnetic monopole configurations also correspond to a fairly simple,
constant curvature space -- $S_2\times\IR$ --
in our geometric formulation of the theory. Moreover, for
spherical configurations, we have studied particular terms in
the electric field energy in the presence of
heavy point sources that lead to the main contribution to the potential
for these sources. We have also identified the manner in which
exact gauge symmetry is maintained locally in the presence of half-integral
spin sources, whereby one must construct half-integer spin
representations from the gauge field variables in order to construct
total angular momentum zero from the addition to the color sources.

Our calculations are by no means complete,
and a number of important
issues must still be considered: for
instance, we have not studied
the Jacobian determinant appearing in the
measure after the change of
variables, $\det |\d A/\d u|$, and we have
not considered the
effects of renormalization. We also expect
infrared effects to appear
once noncompact geometrical configurations
are considered, and these
must be properly treated. Such issues would
form an integral part of
a more detailed computation of, for instance,
the potential energy between two
static color sources. A more detailed study
of the $S_2\times\IR$ geometry would also be
of interest. All such computations are part of
our plans for future work. \medskip

{\bf Acknowledgments}\bigskip

We thank R.R.~Khuri for participation in
the early stages of this work, and D.Z.~Freedman
and R.~Myers for discussions on geometry. KJ would also like to thank
his colleagues at the CTP and in particular
S. Levit for valuable discussions.

\listrefs
\end